\documentclass[twocolumn]{aastex62}
\usepackage{graphicx,graphics,amsmath}
\usepackage{natbib}
\usepackage{color}
\usepackage[normalem]{ulem}
\newcommand{\Ro}{{\rm Ro}}

\shorttitle{cycle period and parity of solar-type stars}
\shortauthors{Hazra et al.}

\begin{document}

\title{Exploring cycle period and parity of stellar magnetic activity with dynamo modeling}

\author{Gopal Hazra}
\affiliation{School of Space and Environment, Beihang University, Beijing, China}
\affiliation{School of Physics, Trinity College Dublin,  the University of Dublin, Dublin-2, Ireland}
\author{Jie Jiang}
\affiliation{School of Space and Environment, Beihang University, Beijing, China}

\author{Bidya Binay Karak}
\affiliation{Department of Physics, Indian Institute of Technology (Banaras Hindu University), Varanasi, India}

\author{Leonid Kitchatinov}
\affiliation{Institute for Solar-Terrestrial Physics, Lermontov Str. 126A, 664033, Irkutsk, Russia}
\affiliation{Pulkovo Astronomical Observatory, St. Petersburg, 196140, Russia}

\correspondingauthor{Jie Jiang}
\email{jiejiang@buaa.edu.cn}



\begin{abstract}
Observations of chromospheric and coronal emissions from various solar-type stars show that the stellar magnetic activity varies with the rotation rates of the stars. The faster the star rotates, its magnetic activity gets stronger but activity cycle period does not show a straightforward variation with the rotation rate. For slowly rotating stars, the cycle period decreases with the increase of rotation rate, while for the fast rotators dependency of cycle period on rotation is presently quite complicated. We aim to provide an explanation of these observational trends of stellar magnetic activity using a dynamo model. We construct a theoretical dynamo model for stars of mass 1 $M_\odot$ based on the  kinematic flux transport dynamo model including radial pumping near the surface of the stars. The inclusion of this near surface downward radial pumping is found to be necessary to match the observed surface magnetic field in case of the Sun. The main ingredients of our dynamo model, meridional circulation and differential rotation for stars are obtained from a mean-field hydrodynamic model. Our model shows a decrease of cycle period with increasing rotation rate in the slowly rotating regime and a slight increase of cycle period with rotation rate for the rapid rotators.  The strength of the magnetic field is found to be increasing as the rotation rate of the star increases. We also find that the parity of the stellar magnetic field changes with rotation. According to our model, the parity flips to quadrupolar from dipolar if the rotation period of the star is less than 17 days.      
\end{abstract}

\keywords{stars: activity, solar-type --- stars: magnetic field, rotation}

\section{Introduction}
Many solar-type stars with an outer convection zone (CZ) show cyclic magnetic behavior like the 11-year cycle of the Sun. Unlike the Sun for which long-term photospheric measurements of the magnetic field and its proxies are available, however, stellar observations are limited. It has been realized that the magnetic (non-thermal) heating in the chromosphere causes an emission in the core of Ca II H \& K lines. This H \& K emission is shown to have a strong correlation with the magnetic flux, as realized in the Sun \citep{Skumanich75,Schrijver92}. Therefore, the H \& K emission flux is taken as a measure of the stellar magnetic activity. While the magnetic activity, in general, is expected to correlate with the rotation rates of the stars, \citet{Noyes84a} showed that the activity correlates better with the Rossby number (\Ro)\ which is a ratio of the rotation period to the convective turnover time. As seen in Figure 8 of \citet{Noyes84a}, the activity first increases rapidly and then the activity increases very slowly or even seems to be independent of \Ro. Combining data of coronal X-ray emission, which is also taken as a proxy of the stellar magnetic field, of more than 800 solar-type stars, \citet{wright11,WD16} find similar rotation-activity relation as seen in the Ca II H \& K data. Like the magnetic activity, the magnetic cycle period also depends on the rotation rate of the star. By analyzing Ca II H \& K data of the homogeneous sample of older slowly rotating stars, \citet{Noyes84b} found that the cycle period ($P_{\rm cyc}$) increases as the rotation period ($P_{\rm rot}$) of star increases. Later many other studies \citep{SB99, Saar2002, Bohm07} also found the similar results that the cycle period increases with increasing rotation period across the different activity branches of stars (e.g., see Figure~1 of \citet{Bohm07}; the cycle period increases with increasing rotation period along inactive and active branches). This increasing trend of cycle period with the rotation period is so called $P_{\rm cyc}$--$P_{\rm rot}$ relation.  Recently, \citet{BoroSaikia18} have anaylized data of 4454 cool stars from various surveys and pointed out that not all stars in their sample follow this relation. Using their robust period detection algorithm, they classified total sample of stars in three different segments: first, the stars with well defined cycles; second, stars with multiple cycle and third, stars with unconfirmed cycle. The stars with well identified cycle periods show an increasing trend of cycle period with increasing rotation period, and most of them are slow rotators belonging to the inactive branch. However, uncertainties lie in the stars of fast-rotating branch, which show the multiple chaotic cycles. This result is also supported by \citet{Olspert18} who have done an individual probabilistic analysis of Ca II H \& K data. Another important property of stellar magnetic activity is the global parity of the magnetic field, which determines the angular momentum loss due to magnetized stellar wind \citep{Reville15}. Although the solar magnetic field is dipolar \citep{Hale19, Stenflo88, DBH12}, we do not have the information about the magnetic parity of other stars.   

Efforts to understand the observed empirical activity-rotation relation and $P_{\rm cyc}$--$P_{\rm rot}$ relation started early using mean-field dynamo models \citep{DR82, Robinson82, Noyes84a, Noyes84b}. In recent years, Babcock-Leighton (BL) type kinematic dynamos aka flux transport dynamo models have become popular models for the solar cycle due to their success in reproducing many aspects of solar magnetic cycle \citep{Bab61, Leighton69, CSD95, Durney95, DC99}; also see reviews \citet{Charbonneau10, Chou11, Karakreview14}. These models are kinematic in the sense that velocity fields are provided from observations and the non-linear feedback due to Lorentz force is not considered in general. The mean flows e.g., differential rotation and meridional circulation play most important role in amplifying and transporting the magnetic fields. Two major components of this model, namely, (i) the generation of toroidal field through the differential rotation---$\Omega$ effect and (ii) the generation of the poloidal field from the decay and dispersal of tilted bipolar active regions---BL process, are observationally supported \citep{Das10,KO11a,Muno13,Priy14,CS15}. In the BL process, the tilt of the bipolar active region is crucial in determining the strength of the poloidal field \citep{Jiang14, Jiang15}. This tilt is believed to be caused by the Coriolis force acting on the rising flux tube in the CZ \citep{Dsilva93}. Thus, it is expected to increase with the rotation rate of the star. Thereby, the magnetic field in the BL dynamo model also should increase with the rotation rate. 

The main difficulty in extrapolating the flux transport model for the stellar case is less availability of the observed mean flows. For the Sun, we have good data of differential rotation and some data of meridional circulation from the helioseismology, and that is the reason why observational data driven models are in very close proximity to explain various observational features of the Sun. Although we have some evidence of differential rotation at the surface of stars \citep{Barnes05, Berdyugina05, Strass09}, the detailed information throughout their CZs is not available. For stellar meridional circulation, we have almost no available data. The detailed information of mean flows inside the stellar CZ is necessary to construct a realistic dynamo model of stars. Therefore, we have to rely on the theoretical analysis to get differential rotation and meridional circulation in stars. One way to get the mean flows for solar-type stars is by solving jointly the mean field equation of motion with heat transport equation \citep{KR95, KS01, Rempel05, Hotta11, KO11b}. Another way is to solve the full Navier--Stokes equation including heat transport equation directly, which is done in global hydrodynamic and magnetohydrodynamic simulations \citep[e.g.,][]{Miesch05, Brown08, Racine11,kapyla16, Karak18}. The mean-field model of \citet{KO11b} calculates the differential rotation of the main sequence dwarfs having different masses and different rotation periods. The meridional circulation automatically comes out in this model, as a consequence of angular momentum balance. When this model is implemented for solar-type stars with solar rotation period, it gives rise the differential rotation that is very close to helioseismology result. This model also gives a single cell meridional circulation encompassing whole CZ with a poleward flow at the surface and an equatorward flow  near the bottom of the CZ. We have used this model to obtain the mean-flows of solar-type stars to incorporate them in all of our dynamo calculations.      

Some of the previous attempts \citep{Jouve10, DC13}, who extrapolated the flux transport dynamo models for the stellar case, have used the mean flows computed from the 3D global hydrodynamic simulations of the stars \citep{Brown08, Racine11, Guerrero13}. \citet{KKC14} have used the mean flows from the mean-field hydrodynamic model of \citet{KO11b}. The activity-rotation relation has been demonstrated in these previous flux transport dynamo models for the solar-type stars \citep{Jouve10,KKC14,KO15}. {However, these models were not able to reproduce decrease trend of cycle period with the increase of rotation rate as seen in the observations of slowly rotating stars}. In the flux transport dynamo model, the cycle period is inversely proportional to the speed of the meridional circulation \citep{DC99, Karak10}. The meridional circulation, on the other hand, decreases with the increase of the rotation rate (as seen in the previous studies that with the increase of rotation rate the energy in the azimuthal motion increases rapidly and decreases the energy in the meridional motion; \citet{Miesch05, Brown08, Kar15}). Thus these models give longer cycle periods for highly rotating stars in contrast to the observations. {In summary, the flux transport dynamo although reproduced the observed stellar activity--rotation qualitatively, failed to reproduce the correct $P_{\rm cyc}$--$P_{\rm rot}$ relation, at least for the slowly rotating stars  \citep[see][for detailed discussion]{Brun14, Chou17}}. The scenario becomes different if turbulent pumping is included in the model \citep{DoCao11}. {Recent global convective simulations also were not able to get the correct trend of cycle period and sometimes get the opposite trend, specifically for the slow rotators \citep{Warnecke18,Strugarek17}.}      

The pumping is unavoidable in a stratified stellar CZ due to the
topological asymmetric convective flow \citep{Tobias01, Kapyla06, MH11}. 
Previous studies have shown that pumping is very important in transporting 
the poloidal field from the surface to the deeper CZ \citep{Guerrero08, KN12}.
\citet{Cameron12} and \citet{Jiang13} showed that a downward radial pumping in the near-surface shear layer 
is essential in the flux transport dynamo model to match the results with the properties of the observed surface magnetic field. 
\citet{KC16} have shown that the pumping increases the dynamo efficiency by suppressing the diffusion of the magnetic field 
across the surface. Thus pumping helps to produce 11-year magnetic cycle even at much high diffusivity in the CZ and also helps the dynamo to recover from Maunder-like extended grand minima of weaker activity \citep{KM17,KM18}. 

The aim of our paper is to explore the $P_{\rm cyc}$--$P_{\rm rot}$ relation using a flux transport dynamo model with added downward radial turbulent pumping.
We shall show that the behavior of the stellar magnetic cycles with the radial pumping will be different than that obtained in previous flux transport dynamo models. {With this model, we shall show that the $P_{\rm cyc}$ decreases with the decreasing $P_{\rm rot}$ of the stars and after a certain rotation period, it starts to increase. we will compare these results with the available stellar observations}.
We shall also predict the parity of the magnetic field in rapidly rotating stars.
We shall show that when the rotation rate is faster than a certain value, the parity 
becomes quadrupolar instead of dipolar which is the dominating parity of the solar magnetic field.
In the next section, we discuss our model and model parameters in detail. In Section 3 we present and discuss our results. Finally, our conclusions are summarized in Section 4. 
      
\section{Model}\label{sec:model}
We adopt the flux transport dynamo model to construct a dynamo model for the solar-type stars. In this model, total magnetic field is assumed to be axisymmetric and consists of toroidal and poloidal components, which can be written as  

\begin{equation}\
{\bf B}_{tot}(r,\theta,t) = B(r,\theta,t){\bf \hat{e}_\phi} + \nabla \times {A(r,\theta,t){\bf \hat{e}_\phi}}. 
\end{equation}        

The following equations govern the behavior of the poloidal field $({\bf B}_p = \nabla \times A(r,\theta,t){\bf \hat{e_\phi}})$ and toroidal field (${B}$) with time
\begin{equation}
\frac{\partial A}{\partial t} + \frac{1}{s}({\bf v}.\nabla)(s A)
= \eta_{t} \left( \nabla^2 - \frac{1}{s^2} \right) A + S_{BL}(r,\theta;B),
\label{eqA}
\end{equation}
\begin{eqnarray}
\frac{\partial B}{\partial t}
+ \frac{1}{r} \left[ \frac{\partial}{\partial r}
(r v_r B) + \frac{\partial}{\partial \theta}(v_{\theta} B) \right]
= \eta_{t} \left( \nabla^2 - \frac{1}{s^2} \right) B \nonumber \\
+ s({\bf B}_p.\nabla)\Omega + \frac{1}{r}\frac{d\eta_t}{dr}\frac{\partial{B}}{\partial{r}},~~~
\end{eqnarray}\\

where $s$ = $r \sin\theta$, $S_{BL}$ is the source function which captures the effect of BL mechanism on the surface of the stars and $\eta_t$ is the turbulent diffusivity and other terms are in usual notation. We want to solve these dynamo equations for different solar-type stars with velocity profiles obtained from the mean field hydrodynamic models of flows. All the model parameters that we have used in our model to understand the stellar magnetic activity are given in details below. We want to point out that our model is somewhat different than the previous model used by \citet{KKC14}. First of all, a radial downward turbulent pumping is used in our model near the surface of the stars and the turbulent diffusivity is higher with respect to the profile that \citet{KKC14} used. We kept the boundary condition for magnetic field completely radial at the surface. The radial pumping near surface and vertical field boundary conditions are required for the results of flux transport dynamo models to match with the observation more closely \citep{Cameron12, Jiang13}. The downward pumping profile mainly contributes in the advection of the poloidal field downwards and less diffusion of fields through the surface.      

\subsection{Meridional Circulation and Differential Rotation of Stars}
\begin{figure}[!htbp] 

\includegraphics[width = 0.4\textwidth]{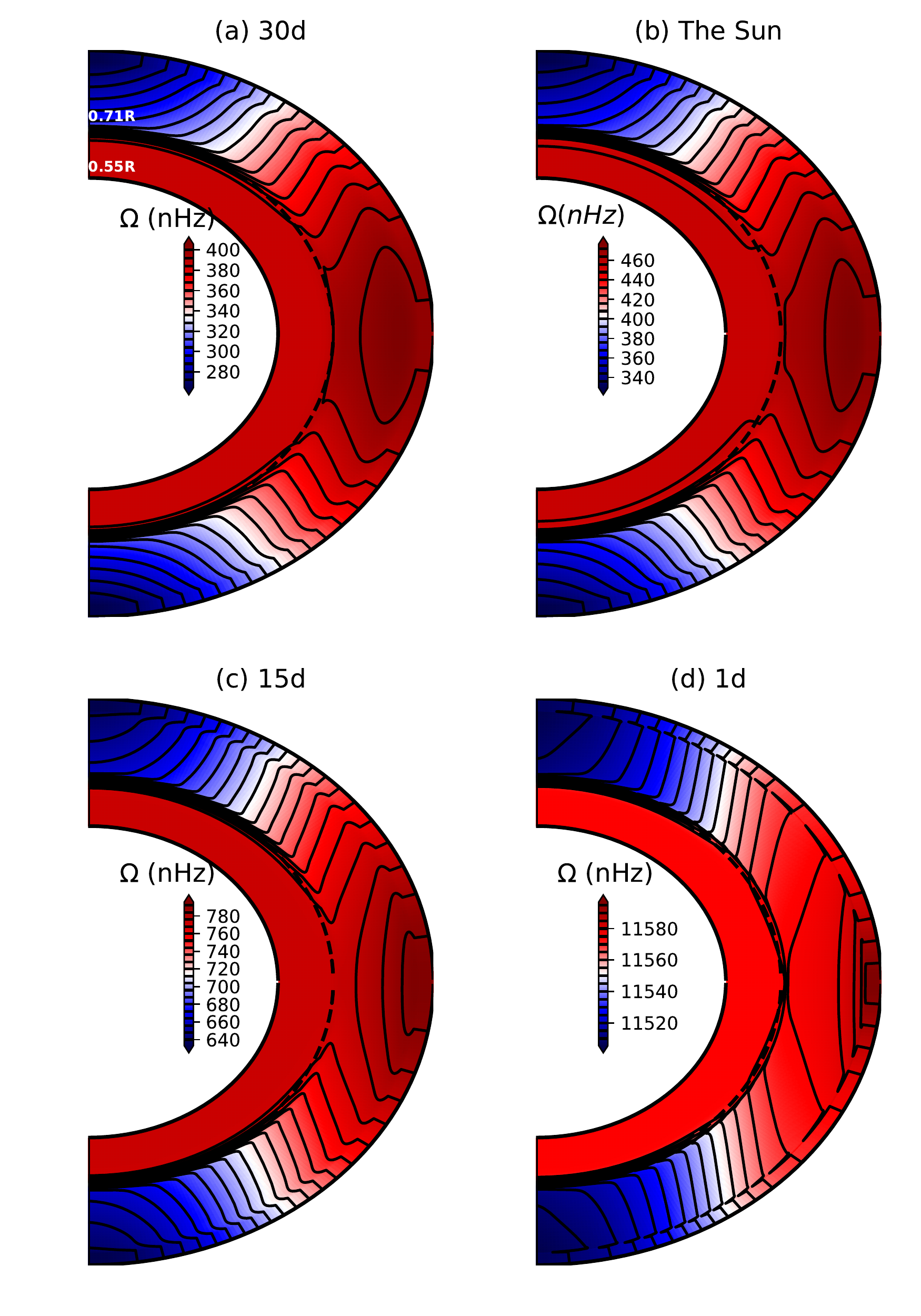}
\caption{\label{fig:diff_rot}Differential rotation computed from the \citet{KO11b} model for stars with rotation period of (a) 30 days, (b) the solar period, (c) 15 days and (d) 1 day}
\end{figure}

\begin{figure}[!htbp] 
 \includegraphics[width = 0.4\textwidth]{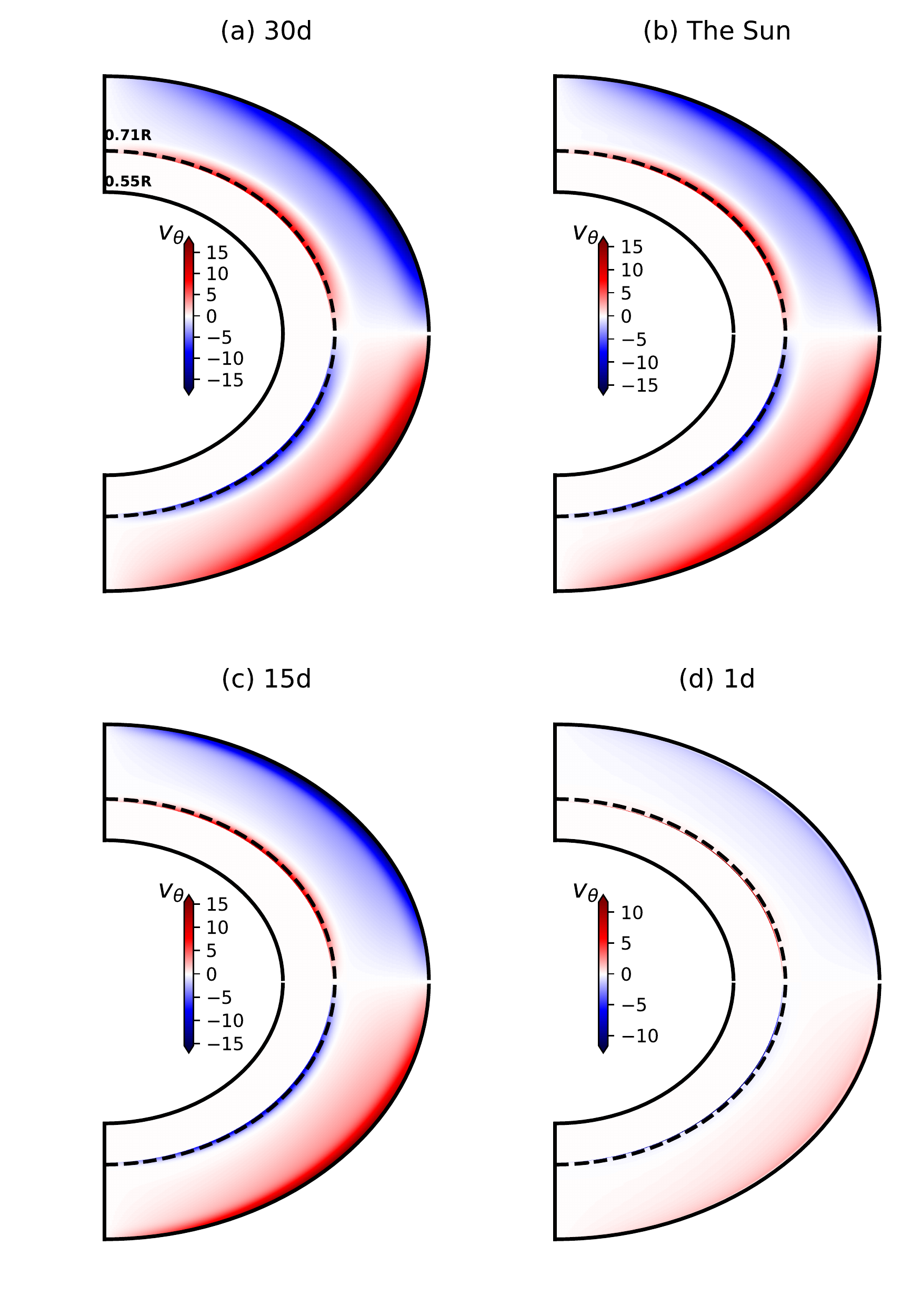}
\caption{\label{fig:mc} Same as Figure ~\ref{fig:diff_rot}, but for the latitudinal component of meridional circulation. All units are in m~s$^{-1}$}
\end{figure}

We have used the same meridional circulation and differential rotation as used by \citet{KKC14} from the model of \citet{KO11b}. Three joint equations for angular velocity, meridional flow and entropy are solved in the CZ of a solar-type star to get the mean flows in this model. The computations give solar-type differential rotation with equator rotating faster than poles for all the stars that we consider in present calculations. The differential profiles for some of the stars with rotation periods of 30 days, the Sun, 15 days and 1 day are given in Figure~\ref{fig:diff_rot}. Meridional circulation for those stars are also shown in Figure~\ref{fig:mc}. The meridional circulations consist of a single cell with a poleward flow at the surface and an equatorward flow at the bottom of the CZ. For all stars, meridional circulation has a penetration depth of $0.7R_{\star}$. Also note that as rotation rate of a star increases, the meridional circulation becomes more confined near the boundaries of the CZ; see Figure 4 of \citet{KKC14}. 

\begin{figure}[!htbp]
\includegraphics[width = 0.5\textwidth]{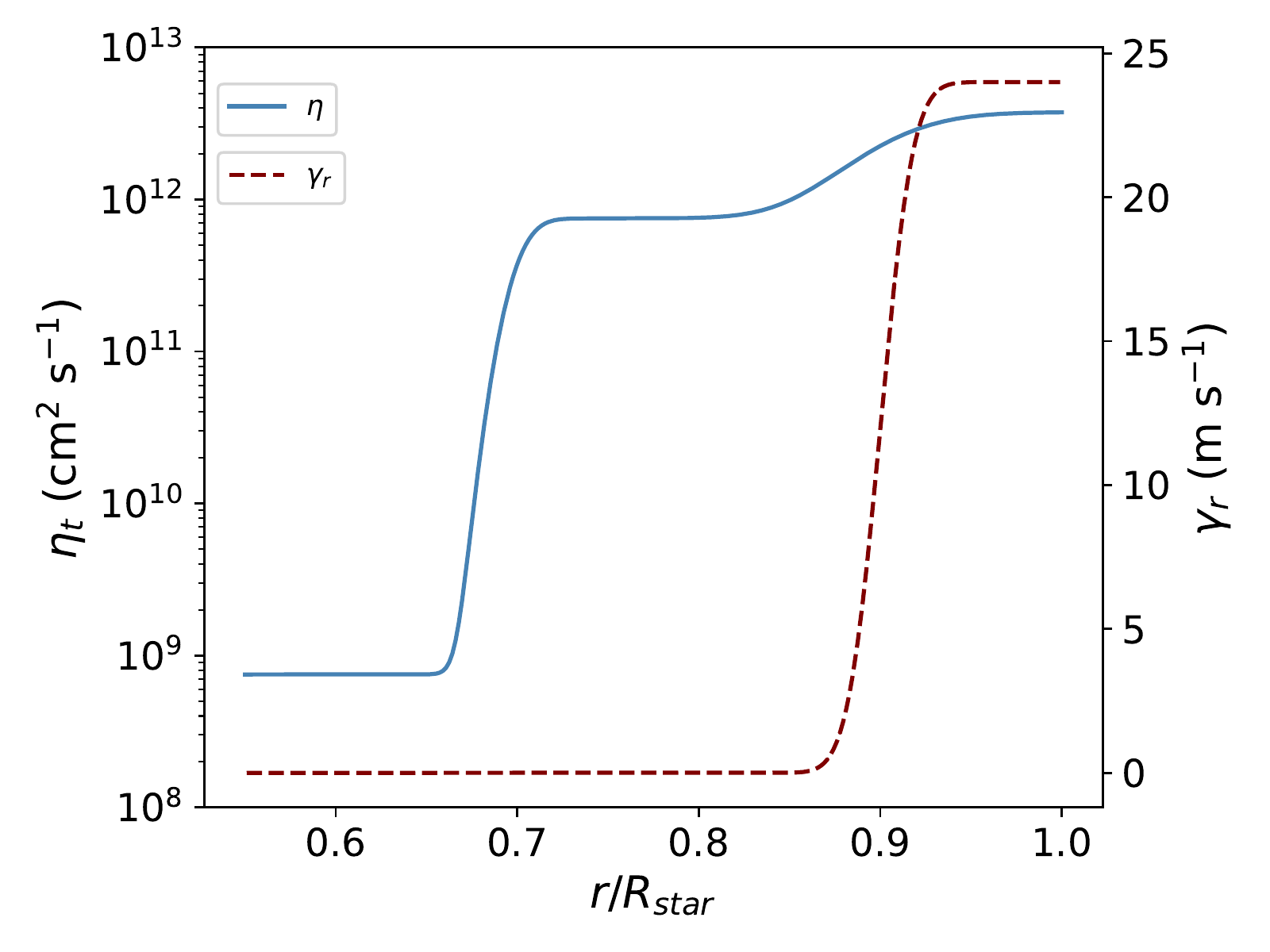}
\caption{\label{fig:eta}Blue solid line shows the turbulent diffusivity used in our simulations and dark red dashed line shows the downward radial pumping near the surface (right y axis shows the amplitude of the pumping).}
\end{figure}

\begin{figure}[!htbp]
\includegraphics[width = 0.52\textwidth]{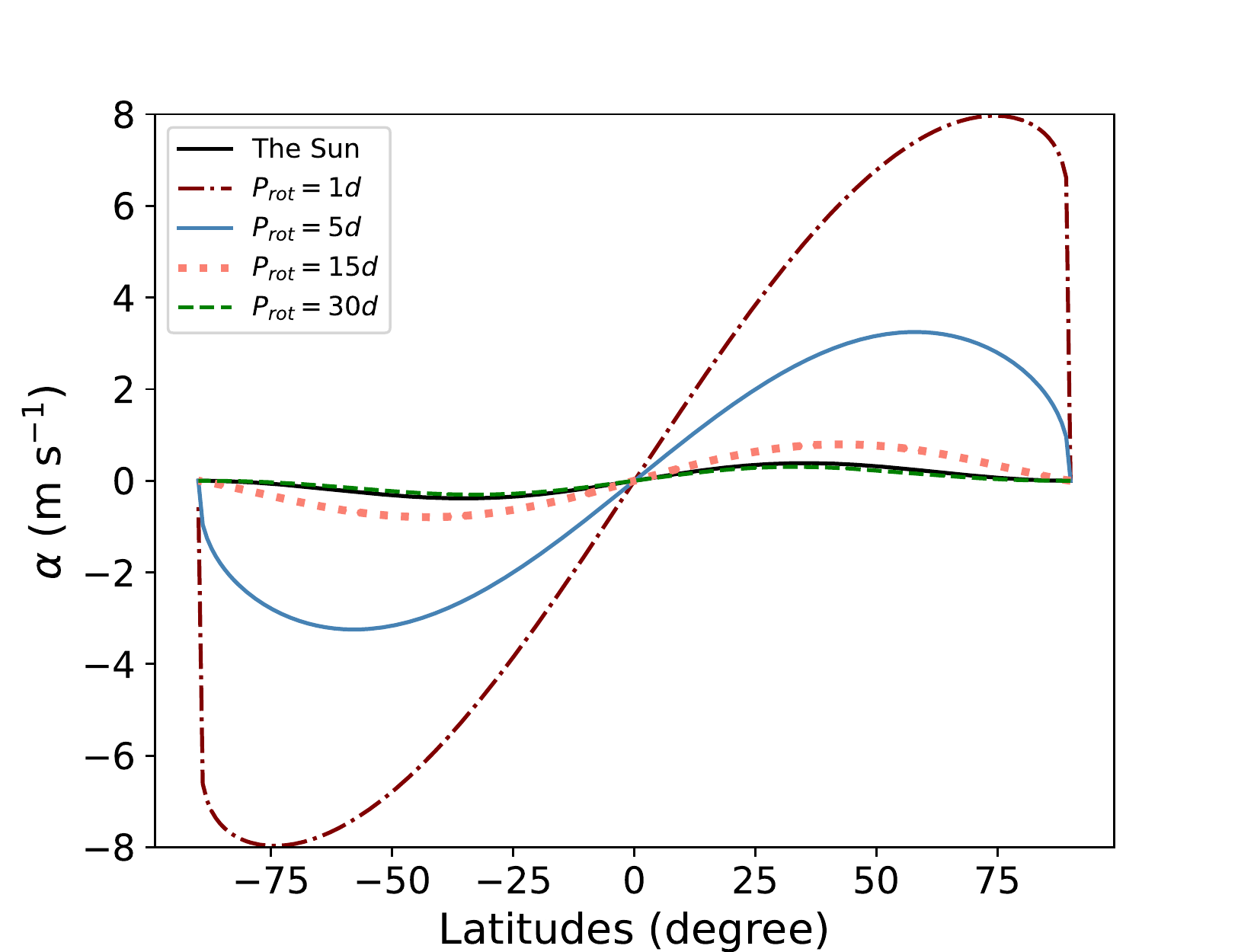}
\caption{\label{fig:new_alpha}The latitudinal variation of $\alpha$-profile for stars with different rotation periods for Case II (Equation~(\ref{case2})). The solid black line shows the standard solar case. As a star rotates faster than the Sun, the corresponding $\alpha$ becomes larger and more concentrated towards high latitudes. Dashdot brown line, blue solid line, dotted orange line and dashed green line show the $\alpha$-profile for the stars rotating with rotation periods of 1, 5, 15 and 30 days respectively.}
\end{figure}

\subsection{Turbulent Diffusivity and Turbulent Pumping}
In most of the flux transport dynamo models, 
an order of magnitude less turbulent diffusivity than what mixing length theory estimate has been 
used \citep[see Figure 4 of][]{KC16}. Here
we use significantly high diffusivity which is close to mixing length estimation
and observationally motivated value of \citet{CS16}. 
The magnetic diffusivity which we have used for all our simulations is  

\begin{eqnarray}\label{eq:eta}
\eta_t = \eta_c + \frac{\eta_{mid}}{2}\left[1 + {\rm erf}\left(2\frac{r-0.7R_{\star}}{0.03R_{\star}}\right)\right] \nonumber \\
~~~+ \frac{\eta_{top}}{2}\left[1+{\rm erf}\left(\frac{r-0.90R_{\star}}{0.05R_{\star}}\right)\right] 
\end{eqnarray} 

where, $\eta_c = 7.5 \times 10^9$ cm$^{2}$~sec$^{-1}$, $\eta_{mid} = 7.5 \times 10^{11}$ cm$^{2}$~sec$^{-1}$ and $\eta_{top} = 3 \times 10^{12}$ cm$^{2}$~sec$^{-1}$.

Following \citet{Cameron12,KC16,KM18}, we use following radial pumping profile near surface. 
\begin{equation}\label{eq:pmp}
\gamma_r = -\frac{\gamma_0}{2}\left[1 + {\rm erf}\left(\frac{r-0.9R_{\star}}{0.02R_{\star}}\right)\right],
\end{equation}
where $\gamma_0$ sets up the amplitude of the magnetic pumping which is fixed to 24 m~s$^{-1}$ in all the simulations, unless otherwise mentioned. Note that choice of $\gamma_0$ value is somewhat arbitrary at present \citep{Cameron12,KC16} and we choose $\gamma_0 = 24$ m~s$^{-1}$ based on the fact that for the solar case, it gives results that are very close to observations. The profiles of $\eta_t$ and $\gamma_r$ are shown in Figure~\ref{fig:eta}. {To be more specific, the idea of incorporating turbulent pumping is to suppress the diffusive decay of the magnetic field through the surface of the stars, and to complete this task we need the near surface magnetic Reynolds number $R_m = \frac{\lvert{v_p}\rvert L}{\eta} > 1$, which implies velocity $v_p  > \frac{\eta}{L}$. 


Assuming $\eta = \frac{\eta_{top} + \eta_{mid}}{2}$ and L = 2$\pi R_\star$, the amplitude of the pumping ($v_p$) approximately should be at least more than 5 m s$^{-1}$ for $R_m > 1$. \citet{Cameron12} performed a series of simulations with different pumping amplitudes and estimated that a strong downward pumping speed 25 m s$^{-1}$ is needed in order to match the results of flux transport dynamo simulation with the observed solar polar fields. Different radial pumping profiles have been used by different groups to mimic the downward transport of the fields \citep{Guerrero08, KN12, Kitchatinov12}. Many of the profiles were confined inside the whole CZ but we have used a downward pumping profile near surface layers \citep{Cameron12, KC16} as radial pumping is most significant there due to very high density stratification, which is consistent with the numerical simulations\citep{MH11} 
}
\begin{figure*}[!htbp]
\includegraphics[width = 0.95\textwidth]{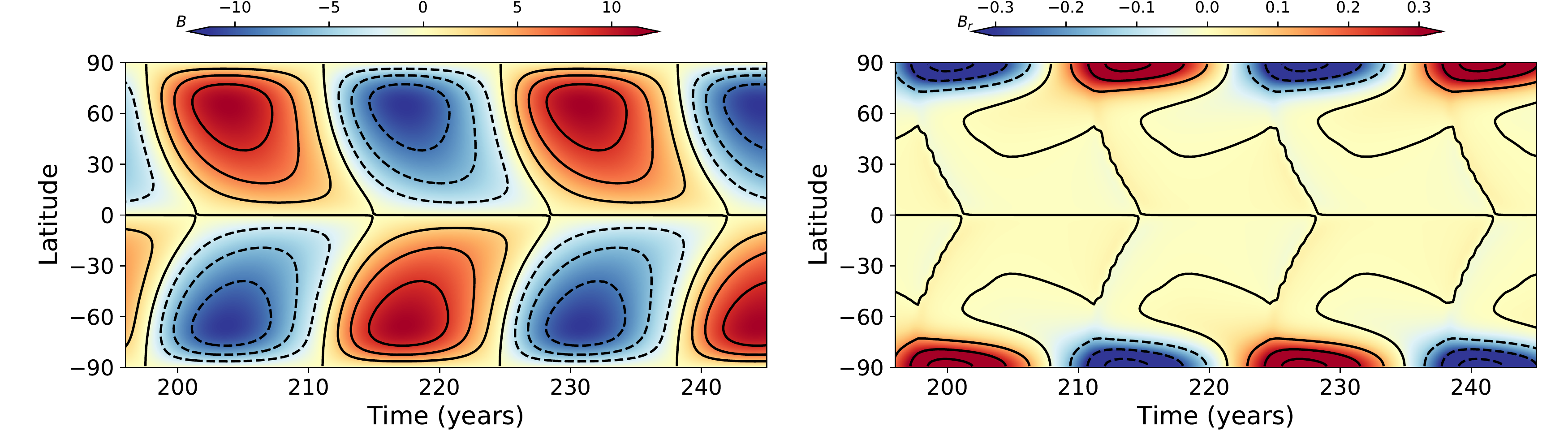}
\caption{\label{fig:bfly_sun}Time-latitude variation of the toroidal field at the bottom of the CZ ($r = 0.71R_\odot$) (left panel) and the surface radial field (right panel). All magnetic fields are given in the unit of $B_0$}
\end{figure*}

\subsection{Babcock-Leighton $\alpha$ }
This is an essential process to generate the poloidal field from the toroidal field in our model and this process is mainly confined near the surface of the Sun. The source term ($S_{BL}$) in Equation~(\ref{eqA}) incorporates this BL process and can be written as
\begin{equation}
S_{\rm{BL}}(r,\theta,t) = \frac{\alpha B(r_t, \theta,t)}{1+[B(r_t,\theta,t)/B_0]^2}
\end{equation}
where $B(r_t,\theta,t)$ is the value of the toroidal field averaged over the tachocline from $r =0.685R_\star$ to $r = 0.715R_\star$. $B_0$ is the quenching field strength and this is the only non-linearity considered in our model. All the magnetic field strengths are given in the units of $B_0$. We have used the following $\alpha$-profile in which we make sure that it should be confined in the upper layers of the CZ. 
\begin{eqnarray}
\alpha = \frac{\alpha(\Omega)f(\theta)}{2}\left[1 + {\rm erf}\left(\frac{r-0.95R_{\star}}{0.01R_{\star}}\right)\right]
\end{eqnarray}
Depending on the rotation rate of a star, the amplitude of the $\alpha$- profile, as well as its latitudinal extent varies. The $\alpha(\Omega)$ determines the strength of the BL mechanism and $f(\theta)$ considers the latitudinal extent of starspot emergence. For rapidly rotating stars, we expect the buoyant rise of the toroidal flux tube along the rotation axis and polar spots would appear in the high latitude regions \citep{SS92, Jeffers02, Waite15, Isik18}. Hence the BL process would be stronger in the high latitude regions of those stars. We have considered two different cases based on two different dependencies of rotation on the strength of the BL process ($\alpha(\Omega)$) and its latitudinal extent ($f(\theta)$).  

First, we have considered the strength of $\alpha$-effect as a simple function of rotation and kept the latitudinal extent same as the standard solar case, 
\paragraph{Case {\rm I}}
\begin{eqnarray}\label{case1}
\alpha(\Omega) = \alpha_0\frac{P_{\odot}}{P_{\rm rot}} ~~~\&~~~ f(\theta) = \rm \cos\theta \sin^2\theta
\end{eqnarray}
Note that the latitudinal emergence of the starspots is taken as a function of $\cos\theta$ (to consider the effect of Coriolis force) and $\sin^2\theta$ is added to suppress the high latitudes emergence of the starspots. 

In the second case, we have used the following dependency of rotation on strength of the BL process motivated by \citet{KO15}.
\paragraph{Case {\rm II}}
\begin{eqnarray}\label{case2}
\alpha(\Omega) = \alpha_0\frac{\sin\left(\alpha_\odot\frac{P_{\odot}}{P_{\rm rot}}\right)}{\sin\alpha_\odot} ~~~\&~~~ f(\theta) = \rm \cos\theta \rm \sin^{2n}\theta
\end{eqnarray}
where $\alpha_\odot$ is average tilt angle observed in case of the Sun and $n =  {P_{\rm rot}}/{P_\odot}$. Here we have considered the rotational dependency on emergence latitudes of starspots, which makes sure that if the star rotates faster, the starspots will have higher latitude emergence. Note that if $P_{\rm rot}$ is equal to the solar rotation period, the $\alpha$ profile converges to the standard solar case. As we do not have exact understanding of how latitudes of starspots emergence vary with the rotation, we just follow a simple profile to invoke the effect of rotation on the latitudinal emergence of starspots. The $\alpha$-profile for this case is shown in Figure~\ref{fig:new_alpha} for a few stars.


\begin{figure*}[!htbp]
\begin{center}
\includegraphics[width = 0.95\textwidth]{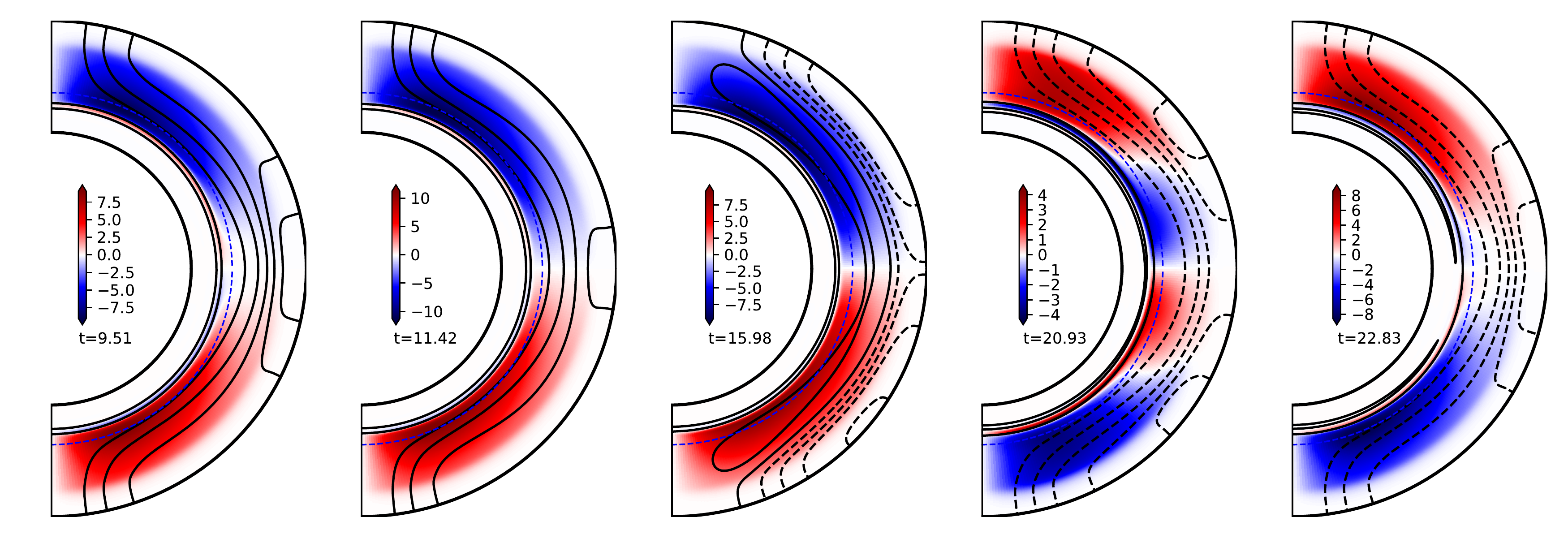}
\caption{\label{fig:pol_tor}The dynamics of the field structure is shown for an entire solar cycle. Five snapshots are plotted at five different times of a solar cycle. The filled contours represent the toroidal field (in the unit of $B_0$), where blue and red colors show the negative and positive polarities respectively. The black contours show the poloidal field lines (solid line represents clockwise and dashed line show the anti-clockwise poloidal fields). }
\end{center}
\end{figure*}

\begin{figure*}[t]
\begin{center}
\includegraphics[width = 1.0\textwidth]{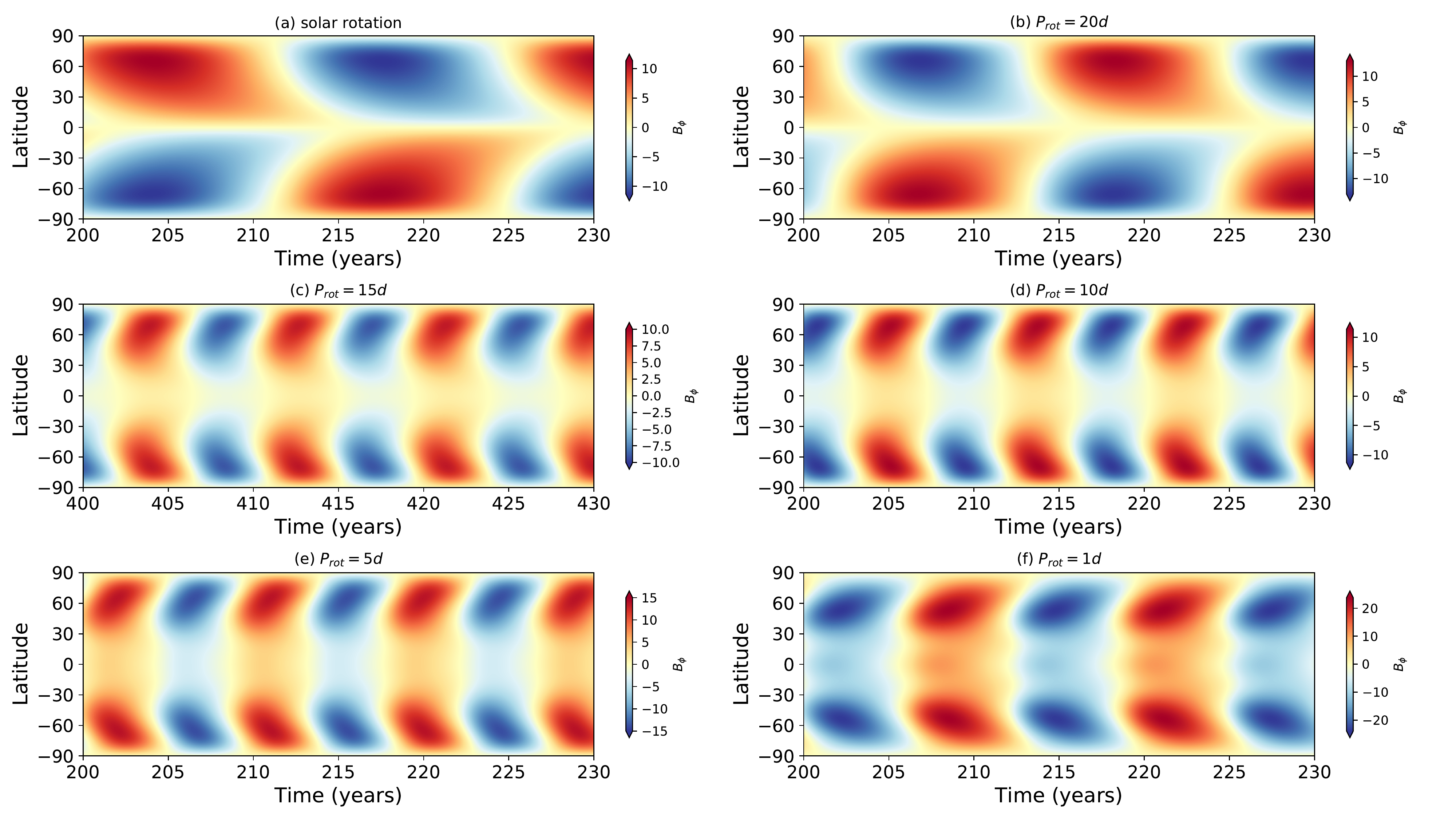}
\caption{\label{fig:bfly1}Time-latitude plots of toroidal field near the bottom of the CZ ($r = 0.71R_\star$) for different stars with rotation period of (a) the solar value, (b) 20 days, (c) 15 days, (d) 10 days, (e) 5 days and (f) 1 day. All plots are for $\alpha$-profile given in Equation~(\ref{case1}).}
\end{center}
\end{figure*}

\section{Results}
We run dynamo simulations for various stars with different rotation periods and mass of 1${M_\odot}$, which have outer CZ and a tachocline. As the parameters, namely, diffusivity and turbulent magnetic pumping, in the present dynamo model differ from the previous models available in the literature, we shall first present the results for the Sun.

\subsection{The Sun}
First, we make sure that with chosen diffusivity, turbulent pumping and BL $\alpha$, a cyclic solar-type solution with 11-year periodicity is reproduced. We find that using the strength of $\alpha_0$ = 4.0 m~s$^{-1}$ and pumping amplitude $\gamma_0 = 24$ m~s$^{-1}$, a cyclic solution and all the basic properties of the solar magnetic field are well reproduced. In Figure~\ref{fig:bfly_sun}, the basic features (e.g., equatorward migration, periodic cycle) of the solar magnetic fields are shown. The toroidal and radial fields are shown in the left and right panels of Figure~\ref{fig:bfly_sun} respectively. The equatorward migration and the correct phase relationship between the poloidal and toroidal fields are evident from the butterfly diagrams. A strong toroidal field appears at relatively high latitudes than expected from the observations of sunspots. High latitude field is a consequence of the strong shear in the tachocline region at high latitudes. The high latitude toroidal flux can partially be avoided by pushing the fields below the tachocline by the downward meridional flow \citep{Nandy02}; however, our meridional flow does not penetrate below the tachocline. The five snapshots of toroidal and poloidal field lines are also shown in Figure~\ref{fig:pol_tor} across an entire solar cycle. Filled contours show the toroidal fields and black solid and dashed contours show the poloidal fields at same five instants of time.

A specific aspect of our solar dynamo solution is that the cycle period is about 13 years. 
This time period is important not because of this number is close to the solar cycle period rather this was not expected
given the value of the diffusivity (Equation~(\ref{fig:eta})) used in our model. It is expected that the dynamo cycle period becomes shorter on increasing the diffusivity \citep{DC99,Hotta10}. In fact, the cycle period cannot be longer than the diffusion time-scale of the magnetic field. That is why all previous flux transport dynamo models were built at much lower diffusivity to get the 11-year period \citep{DC99,Jouve08}. The exceptions were (i) the Surya model \citep{CNC04, Jiang07} in which the poloidal field diffusivity was reasonably high but the toroidal field diffusivity was much lower and (ii) the model of \citet{KO15} in which the diffusivity in the upper convection is high but much lower below $0.75R_\star$. In the present model, we get a longer period at relatively higher diffusivity because of using a downward magnetic pumping. As demonstrated in \citet{Cameron12} and \citet{KC16}, the pumping makes the magnetic field more radial near the surface and suppresses the diffusion of the flux across the surface. The weaker diffusion effectively increases the dynamo efficiency and allows dynamo action at a smaller value of $\alpha_0$.  This causes to increase the cycle period. Essentially what happens is that the poloidal field remains frozen in the CZ for a long time and this allows the shear to produce a strong toroidal field which also does not diffuse across the surface and only diffuses across the equator (see Figure~\ref{fig:pol_tor}). This makes the cycle longer.


\subsection{Stars with Different Rotation Periods}
Now we present results of our simulations for different stars with different rotation periods. 
The mass of these stars is 1M$_\odot$ and the rotation periods are 
1, 2, 3, 5, 7, 10, 12, 15, 17, 20, 25.38 (the solar value) and 30 days. 
Since the rotation has a significant effect on the dynamics of the CZ, 
the mean flows and other turbulent transport coefficients are expected to be 
different for starts with different rotation periods. As discussed in the model section,  
the mean flows that we have used in our model are different for different stars \citep{KO11b}.  
But to avoid complicacy, we consider the same turbulent diffusivity and pumping as that of the Sun (see Equations~(\ref{eq:eta}) and (\ref{eq:pmp})) for all stars. Transport coefficients are also dependent on the magnetic field \citep{Kitchatinov94, karak14} and thus are expected to change with different stars. We however, ignore these in our model.
We now show the results of simulations with $\alpha$ profile as given in Equation~(\ref{case1}).

All the stars in our sample show the systematic periodic variation in their activity cycle. 
Figure~\ref{fig:bfly1} shows the time-latitude diagrams of toroidal magnetic field at $r=0.71R_\star$ 
for six different stars. We believe that this toroidal field 
largely governs the dynamics of the starspots on the stellar surface.  
The slowly rotating stars with rotation period $\ge 17$ days 
show some features which are common in the Sun, particularly the equatorward migration of 
toroidal field and dipolar magnetic fields as shown in Figures~\ref{fig:bfly1}(a) and (b). 
However, for the stars with rotation period of 15 days and less, the equatorward migration 
turned into poleward migration and the parity changed to the qudrupolar.  
The poleward migration is due to the dominant role of the dynamo wave over the meridional flow.
As discussed in Section~\ref{sec:model}, with increasing rotation rate the meridional circulation 
becomes increasingly weaker and the transport of the toroidal field by the 
meridional flow becomes increasingly less effective (Figures \ref{fig:bfly1}(c)--(f)). 
When the rotation period decreased to 15 days, the toroidal band is completely dominated 
by poleward propagating branches. These results are more or less insensitive to our 
chosen BL $\alpha$ profiles. Interestingly, for rapidly rotating stars starspots
are also observed in the higher latitudes. Therefore, the stronger toroidal field (which are believed to
produce the starspots) appeared at higher latitudes in our rapidly rotating stars are consistent with observations. However, recent simulation by \citet{Isik18} based on flux emergence and SFT model shows the equatorward propagation of starspots even in case of fast rotating stars. This is presumably because of the solar-type time-latitude pattern of flux eruption that they have implemented at the base of the CZ in all solar-type stars including the rapidly rotating stars. They have also found that for fast rotating stars, the activity near-equatorial region is activity-empty because of flux-tube rise along the rotation axis. Our calculations find that for fast rotating stars, the toroidal field near the equatorial region is less due to cross equatorial diffusion but its effect is not negligible (see Figure~\ref{fig:pol_tor_1}).   

\begin{figure*}[t]
\begin{center}
\includegraphics[width = 1.0\textwidth]{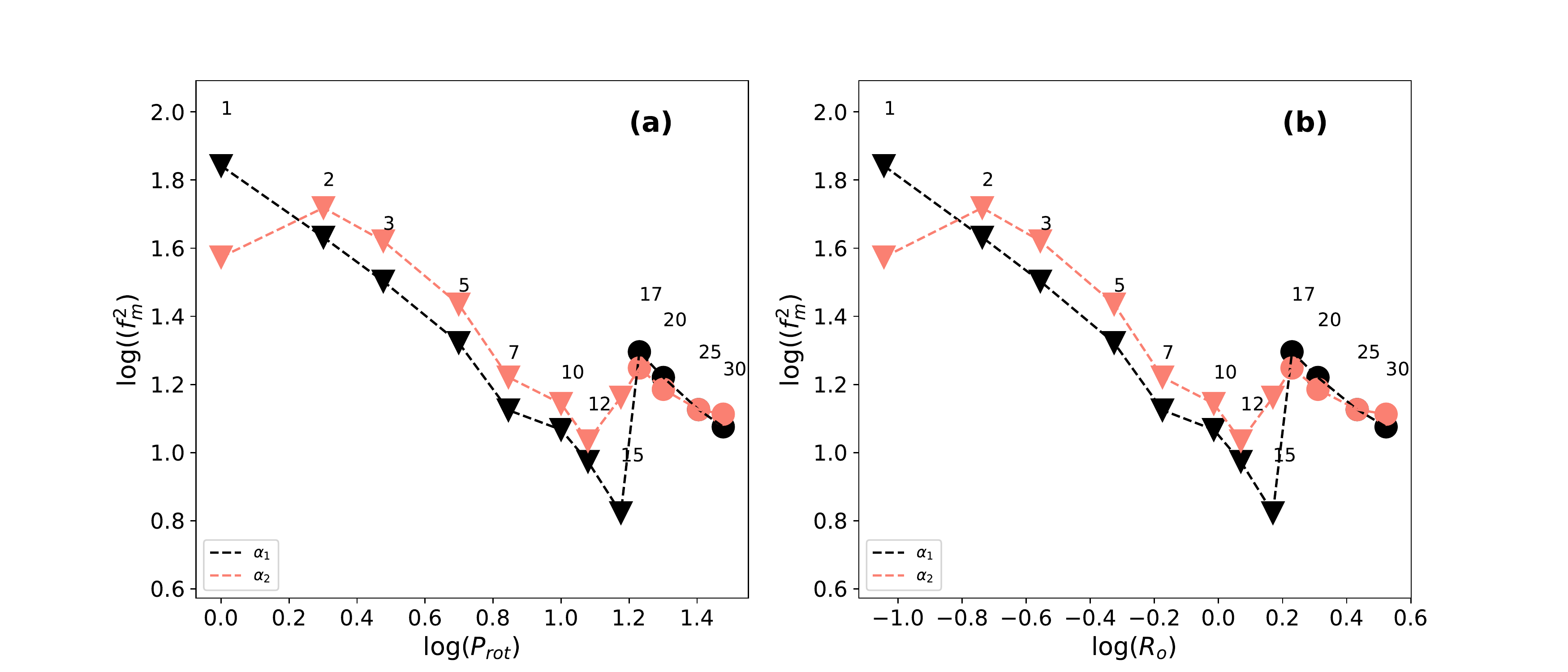}
\caption{\label{fig:max_tor}The peak amplitudes of the toroidal flux for different stars are plotted as a function of (a) rotation period and (b) Rossby number. Two different curves in each plot are for two different $\alpha$-profiles (Equation~(\ref{case1}) and (\ref{case2})). Triangular and circular symbols show stars with quadrupolar and dipolar fields respectively. Numbers on the plot represent the rotation period of each star.}
\end{center}
\end{figure*}

\subsection{Activity-Rotation Relation}
Now we want to see whether our model can explain the observed variations of the Ca II H \& K 
emission with rotation. The amount of Ca II H \& K emission flux from the stellar chromosphere 
is a direct consequence of the magnetic heating in the stellar chromosphere, 
hence it should be related to the total magnetic flux of the stars that is generated inside 
their CZs. We first measure the total amount of toroidal flux inside the stellar CZ  
as $fB_0R_{\star}^2$, where $B_0$ is the quenching field strength (same for all of our calculations) 
and then we consider the value $f$ as a measure of total toroidal flux for all of the stars.
As discussed in \citet{KKC14}, the nonlinear quenching in $\alpha$ tries to limit the magnetic field in the model
around $B_0$. Thus the stellar magnetic flux should be measured with respect to $B_0 R_{\star}^2$. 
Obviously, if we plot  $f$ as a function of time, we find an oscillatory behavior and
its value during the cycle maxima would be maximum and we call it $f_m$.
As the rotation of a star increases, we expect more field to be generated inside stellar CZ 
and that would lead to more Ca II H/K emission. Since the production of the emission generally involves magnetic reconnection of one flux system with another, we naively assume the emission to go as the square of the magnetic flux i.e., $f_m^2$ following \citet{KKC14}. {\citet{Vidotto14} found the value of 1.8 $\pm$ 0.2 for the power index in the dependence of the X-ray luminosity on the large-scale (unsigned) magnetic flux (cf. very short Sect. 3.1.4 and Fig.5 in their paper).} As we have used two particular profiles of BL $\alpha$ depending on the rotational dependency as given in Case I (Equation~(\ref{case1})) and Case II (Equation~(\ref{case2})), we present results using both of them.
The square of maximum toroidal flux amplitudes ($f_m^2$) is plotted as a function of rotation period, which is shown in Figure~\ref{fig:max_tor}(a) with $\alpha$ for Case I and $\alpha$ for Case II. In Figure~\ref{fig:max_tor}(b), we plot the same square of maximum amplitude of the toroidal flux but with Rossby number instead of rotation period. Since our sample points spectrally belong to G type of stars, both of the Figures~\ref{fig:max_tor}(a) and \ref{fig:max_tor}(b) show similar trend. {As we see with both $\alpha$-profiles, the toroidal flux keeps on increasing with increasing rotation rate but for the stars with rotation period less than 17 days (when parity flips from dipolar to quadrupolar), the total magnetic flux encounters diminution and after that it again starts to rise.  For the $\alpha$ profile  chosen in Case II, we find a dip for the fast rotators with rotation period of 1 day, which is not found for rotation profile of Case I. 

The amplitude of the toroidal field which determines the cycle strength kept on increasing with increasing strength of the BL $\alpha$. In both cases, the strength of $\alpha$ always increases with rotation and the strength of the toroidal flux also kept on increasing. When parity changes from dipolar to quadrupolar for stars with rotation period less than 17 days, the cross equatorial diffusion of toroidal flux across the hemispheres becomes more and as a result we find a diminution in the toroidal field strength (see Figure~\ref{fig:max_tor}).  A slight dip which is also found for the toroidal flux during fast rotating stars for $\alpha$ profile chosen in Case II is because of the inclusion of tilt angle saturation in the specified $\alpha$ profile.  For the Case I (Equation~(\ref{case1})) where rotational dependency on $\alpha$ is chosen as $\alpha(\Omega) = \alpha_0\frac{P_{\odot}}{P_{\rm rot}}$, the strength of $\alpha$ always increases with rotation and the strength of the toroidal flux also kept on increasing (black dashed line of Figures~\ref{fig:max_tor}(a) and (b)). This is in accordance with the findings of \citet{KKC14} with same rotational dependency on $\alpha$. In Case II, we have chosen the saturation in the BL $\alpha$-profile (Equation~(\ref{case2})) motivated from observations. In principle, the tilt-angle of bipolar magnetic regions would increase with the rotation rate of a star. Also, we expect high latitudes emergence of the bipolar regions i.e., the radial rise of flux tubes for rapid rotation. These all properties of flux tube emergence are incorporated in the profile that we use for BL $\alpha$ as in Case II (Equation~(\ref{case2})). Incorporating tilt-angle saturation makes the poloidal fields to be saturated for rapid rotators, and allowing radial rise contributes less flux in the surface \citep{KKC14}, which gives rise the dip in the highly rotating branch of the rotation-activity curve (Figure~\ref{fig:max_tor}). As our model is presently kinematic, we have not considered the Lorentz force feedback on the mean flows of the stars \citep{HC17}. The Lorentz force feedback for fast rotators most likely modulates the mean flows of those stars which in turn modulates the total magnetic field of those stars. 

\subsection{$P_{\rm cyc}$--$P_{\rm rot}$ Relation}
We have also calculated the cycle periods based on the toroidal field reversals for different stars rotating with different rotation rates. The cycle period keeps on varying with increasing rotation rate or decreasing rotation period. In Figure~\ref{fig:rot_period}, we have shown two cases with two $\alpha$-profiles used in our model. For both of them, we find that the cycle period decreases with decreasing rotation period, and for the stars with rotation periods less than 15 days, cycle period increased very marginally and for fast rotators (after rotation periods of 7 days), it started to increase considerably. The different symbols in the plot show the parity of the global magnetic field of the stars. Triangle symbols show the stars with quadrupolar parity and filled circles show the stars with dipolar parity. Note that after the parity flips to quadrupolar for the stars with rotation periods of less than or equal to 15 days, the cycle periods started to increase.
{Overall our results are consistent with the observed $P_{\rm cyc}$--$P_{\rm rot}$ relation. We find a decreasing trend of cycle periods with increasing rotation rate for slowly rotating stars as found by \citet{BoroSaikia18} (See Figure~9 of their paper). For the fast rotating stars, there is no clear observed trend of how cycle period behaves with the rotation rate of the stars and our results provide some crucial evidence.} As mention earlier, in all of our simulations, we have not changed the transport coefficients i.e., turbulent diffusivity and turbulent pumping which may have a significant contribution to settle up the cycle period. So far, we do not have a clear indication that how these transport coefficients will vary with the rotation, and we just keep everything constant with rotation. The blue dashed line in Figure~\ref{fig:rot_period} shows the cycle period as a function of rotation period for the $\alpha$ in Case II (Equation~(\ref{case2})). It shows the similar behavior as the black dashed line that has the same pumping amplitude but different $\alpha$ (Case I, Equation~(\ref{case1})). {Hence $P_{\rm cyc}$--$P_{\rm rot}$ relation is robust under the variation in the chosen profiles of  $\alpha$}.     

The previous efforts \citep{Jouve10,DC13,KKC14} to model the magnetic activity for the solar-type stars were not able to get the correct trend of cycle period with the rotation. {However, \citet{DoCao11} found the decreasing trend of cycle period with increasing rotation rate by incorporating rotation dependent radial and equatorward turbulent pumping inside the whole CZ.} Unlike \citet{DoCao11}, we have incorporated only radial pumping near the surface and have chosen a high magnetic diffusivity inside stellar CZ. In our model, the cycle period is dominated by radial turbulent pumping and turbulent diffusion. For the sample of stars that we have selected, apart from the varying rotation rate, the main parameters which change to affect the cycle period are the meridional flow and the strength of BL $\alpha$. But in the presence of turbulent radial pumping and high turbulent diffusivity, the cycle period becomes less dependent on the meridional flow. When rotation rate increases, the differential rotation or the shear becomes very strong allowing generation of more toroidal fields, which in turn helps to produce more poloidal fields. Since the strength of the $\alpha$ parameter also increases with increasing rotation rate, the fast rotating stars have more poloidal field to reverse the existing toroidal field, which makes cycle shorter \citep{KC16}. This trend is clear from our simulations. As the rotation rate of a star increases, the period of the cycle tends to be shorter as well as the amplitude becomes stronger. For the stars with rotation periods less than 15 days, there is a trend of a slight increase in cycle period with decreasing rotation period. The radial shear becomes weak for those stars and it takes more time to generate the opposite polarity toroidal fields by twisting the poloidal fields, which affects the timescale of polarity reversal and hence the cycle periods.

\begin{figure}[!htbp]
\begin{center}
\includegraphics[width = 0.5\textwidth]{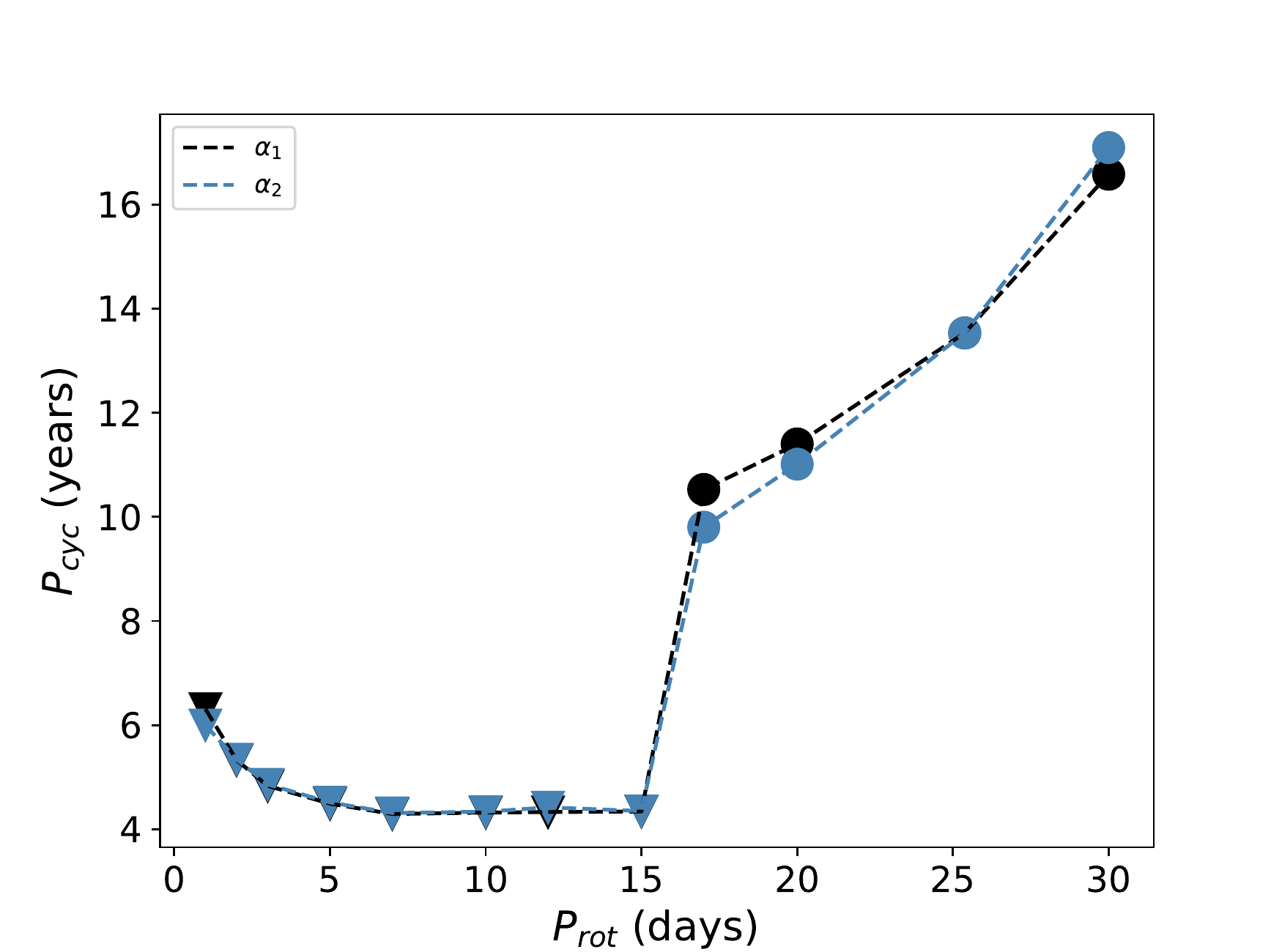}
\caption{\label{fig:rot_period}$P_{\rm rot}$ vs $P_{\rm cyc}$ plot for two different $\alpha$- profiles used in our simulations. Black dashed line shows the case in which $\alpha$ profile is chosen based on Equation~(\ref{case1}). The blue dashed line shows the case in which $\alpha$ profile is based on Equation~(\ref{case2}). Triangular and circular symbols show stars with quadrupolar and dipolar parities respectively.}
\end{center}
\end{figure}

\subsection{Parity}
Another intriguing result that we obtained from our simulations is the change of stellar magnetic field parity with rotation. The parity becomes qudrupolar for the stars with rotation period less than 17 days. We have not run our simulations for many rotation periods to determine the exact rotation period after which the magnetic fields flip to quadrupolar. But qualitatively we find that if the rotation periods are less than 17 days, the stellar magnetic field becomes qudrupolar. The parity of the global magnetic field of stars is determined by various turbulent transport coefficients inside stellar CZ. Whether parity of stellar magnetic field  would be dipolar or quadrupolar is mainly decided by two factors. First, how the toroidal field inside the CZ couples with the opposite hemisphere and second, how efficient is the equatorward mixing of the poloidal field near the surface. To keep the magnetic field anti-symmetric across the equator i.e., to keep dipolar parity, the hemisphere coupling of the toroidal fields should be very less. Usually, the amount of  turbulent diffusivity used in the solar CZ is two orders of magnitude less than its value near the surface and the cross equatorial diffusion of toroidal fluxes are very less, which makes sure that the polarity of the toroidal field in both the hemisphere remains opposite to each other.  Also as BL mechanism is confined near the surface, a good hemisphere mixing is necessary near the equator for leading polarity starspots to cancel out each other. That requires a high value of turbulent diffusion near the surface to maintain the dipolar parity. 

In our present model, we have used high diffusivity value inside the CZ and a turbulent radial pumping near the surface. Turbulent radial pumping near the surface  and radial boundary condition in our model make the field lines radial near the surface and do not allow diffusive decay of fields through the surface.  So in this situation fields can only diffuse through the equator. But for fast rotating stars, the poloidal field does not get enough time to diffuse across the equator because the cycle period becomes shorter as a star rotates faster. Also, for fast rotating stars, the strong poloidal field is generated at very high latitudes that does not get enough time for the diffusive mixing across the equator. As a result, the fast rotating stars turn out to have a quadrupolar polarity. The less diffusion of the poloidal field near the equator makes the new toroidal field symmetric across the equator. Therefore, for rapidly rotating stars (e.g., $P_{\rm rot} = 1$ day), the equatorward cancellation of poloidal field is not that efficient (see poloidal field lines of Figure~\ref{fig:pol_tor_1}), which in turn prefer the quadrupolar mode. Observationally, it is very difficult to quantify the correct parity of the stellar magnetic fields due to instrument limitations, and there are no direct observed results till now about the correct parity of the stars. Our simulations provide very important information in this area about the correct parity of the stars.    
 
\begin{figure*}[!htbp]
\begin{center}
\includegraphics[width = 0.95\textwidth]{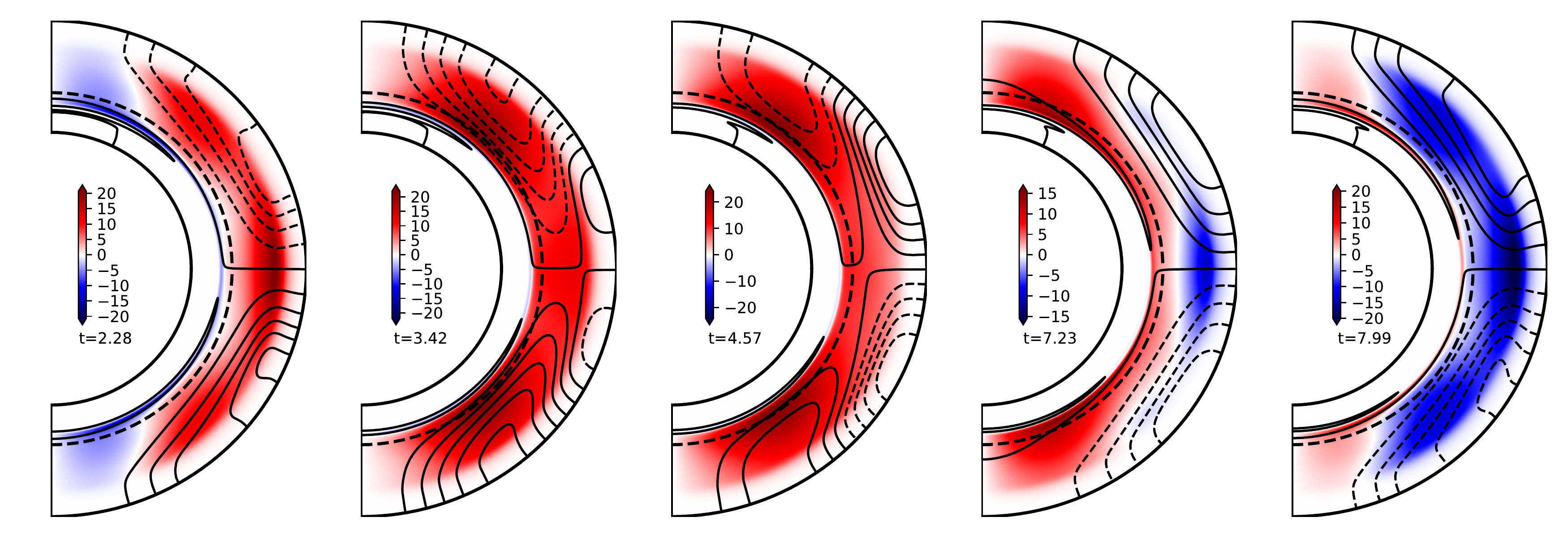}
\caption{\label{fig:pol_tor_1} Same as Figure~\ref{fig:pol_tor} but for the star with rotation period of 1 day.}
\end{center}
\end{figure*}

{In summary, the most important results that we have obtained from our simulations are -- (1) the global magnetic fields of the stars tend to be quadrupolar for rapidly rotating stars, more precisely, in all of our simulations, we get quadrupolar parity of the stars whose rotation periods are less than or equal to 15 days; (2) the cycle period decreases as rotation rate of the star increases and after rotation period of 15 days, there is an increasing trend in cycle period as rotation rate of star increases ; and  (3) an increasing magnetic activity as rotation rate of the star starts to increase.} The first two results are new results and new achievements from our model. 
Previously \citet{DoCao11} was able to reproduce {the decreasing trend of cycle period with increasing rotation rate} but they had to assume a rotation dependent turbulent pumping throughout the CZ, which scales as $\Omega^2$ and their equatorial pumping was strong compared to the radial pumping. Whereas we have assumed only the constant radial pumping near surface layers of the stars, which is more physical and consistent with previous studies \citep{MH11, Cameron12, Jiang13}.

{Both of these results (quadrupolar fields and $P_{\rm cyc}$--$P_{\rm rot}$ relation) are robust under reasonable changes in the parameters adopted in our models. The overall behavior of the results remains same if we vary the amplitude of the turbulent pumping. We run a few cases for all stars considering a range of turbulent pumping amplitudes from 16 -36 m s$^{-1}$  and find that the each case gives similar trend in the  results. However, if the pumping amplitude is reduced below 16 m s$^{-1}$ then the decrease of cycle period with the increase of rotation rate disappears. We have also performed some of the simulation by changing the pumping amplitude with rotation. Notably, the set of simulation in which the turbulent pumping is quenched linearly with the rotation period, also produce similar behaviors. The dynamo saturation for the rapidly rotating stars was explained earlier \citep{Jouve10, KKC14, KO15} but we present it here to support the idea that dynamo saturation might be a result of saturation in the BL mechanism and we find an indication of the dynamo saturation using our $\alpha$ profiles given in Case II. For Case II, we have used the similar profile motivated by \citet[see Equation (9)]{KO15} and we find quite similar results. One of the other possibility of the dynamo saturation might be the Lorentz force feedback on the mean flows, which we could not explore in our model due to its kinematic nature.}

\section{Conclusions}
In this paper, we have utilized a kinematic flux transport dynamo model with BL $\alpha$ as the poloidal
source to explain the features of the magnetic fields and cycles of solar-type stars with mass 1~M$_\odot$.
Our dynamo model is different than the traditional flux transport dynamo models in the sense that our model includes
a significant downward magnetic pumping which mimics the strong asymmetric surface convection \citep{MH11}.
This magnetic pumping suppresses the diffusion of the horizontal magnetic field through the surface and thus makes the behavior of
dynamo different than traditional flux transport dynamos \citep{Cameron12, Jiang13, KC16, KM17, KM18}. In our dynamo model,
the large-scale flows such as differential rotation and meridional circulation 
are taken from the hydrodynamic mean-field models \citep{KO11b} of the corresponding stars.
The back reaction of the magnetic field on the flow has been ignored in this model. In all our simulations, we have used higher diffusivity in the CZ ($\eta_{CZ}$) $7.5 \times 10^{11}$ cm$^2$~s$^{-1}$ than the traditional flux transport dynamo model where they use 5 $\times$ 10$^{10}$ cm$^2$~s$^{-1}$ \citep{DC99, MD14}. Some exceptions are \citet{CNC04, HKC14} who use different diffusivities corresponding to poloidal and toroidal fields. 
 
The $\alpha$ profile is changed in two different ways. In one case, 
only the amplitude of the $\alpha$ profile is scaled up with the rotation rate (Case I; Equation~(\ref{case1})),
while in the other case, the mean sunspot tilt is scaled up with the rotation rate 
and the latitudinal profile of the $\alpha$ is increased with the rotation rate such that
with the increase of the rotation rate the region of the generation of poloidal field
is progressively moved to the higher latitudes (Case II; Equation~(\ref{case2})).
The latter case is more physical because we know that in the BL process, the average tilt ($\lambda$) of BMR
largely determines the poloidal field. Thus BL $\alpha$ should be proportional to $\sin({\lambda})$. 
The BMR tilt is expected to increase with the rotation rate (due to the increase of Coriolis force as suggested in thin flux tube simulations \citep{Dsilva93,Fan93}). Observations also suggest that the starspots appear in higher latitudes \citep{Jeffers02, Marsden06, Waite15} and thus we expect that the region of the poloidal field generation would be more concentrated towards high latitudes for rapidly rotating stars.

{An interesting result of our study is that the cycle period decreases with the increase in the rotation rate of a star for slowly rotating stars and the cycle period starts to increase for a star whose rotation period is shorter than 15 days. 
This result is in agreement to a greater extent with the available observed magnetic activity of solar-type stars \citep{Noyes84b,Suarez16, BoroSaikia18}.}
This result also is in striking contrast to the previous studies based on the traditional flux transport dynamo models \citep{Jouve10, KKC14}.
As the meridional circulation becomes weaker in the rapidly rotating stars, we expect the cycle period to be longer. However, in our model, the downward magnetic pumping does not allow the field to diffuse across the surface
and thus the field can only diffuse across the equator. And as the rotation of the stars started to increase, the poloidal field becomes stronger which takes less time to reverse the toroidal field faster, reducing the cycle period.

The most appealing result of our study is that the magnetic field changes its parity from dipolar to
quadrupolar in the rapidly rotating stars of rotation period less than 17 days. In the rapidly
rotating stars, shorter cycle period (as discussed above) gives less time to diffuse the field.
The less diffusion of the poloidal field across the equator does not allow to connect the poloidal field
which is essential to keep the field dipolar. 
This inefficient mixing of the poloidal field across equator near the surface makes
the magnetic field quadrupolar in rapidly rotating stars of the periods less than about 17 days.
Thus our simulations predict quadrupolar field in rapidly rotating stars.

{Another important result of our study is that the magnetic activity increases with the increase of rotation rate associated a small dip at rotation period of 15 days. In Case II of $\alpha$- profile, we find a slight indication of saturation of magnetic field in agreement with \citet{KO15}. The increase of magnetic activity in our model is in broad agreement with observations \citep{Noyes84a, wright11, Vidotto14, Suarez16}.}

Although our results are robust under some variations in the parameters (e.g., BL $\alpha$ and turbulent pumping), we have some limitations in our study. We have not explored a wide range of the diffusivity profile. The turbulent diffusivity even for the Sun is not well constrained. Theoretical and observational arguments \citep{munoz11, Chae08, CS16, HM18} suggest that in the surface layer it is
at least $10^{12}$ cm$^2$~s$^{-1}$ and in the deeper CZ it is probably less. However, we have limited knowledge of how it varies with the rotation rate and the magnetic field strength in other stars \citep[e.g.,][]{Kit94,karak14}. Therefore, a drastic change in the turbulent diffusivity (and also in the magnetic pumping) may alter our conclusions reported in this study. Our model is kinematic, meaning we have ignored the feedback of magnetic field on the large-scale flows. While for the Sun probably this is not a bad
approximation because the variation in the differential rotation with the magnetic cycle is very tiny. However, this may not be a good assumption in rapidly rotating stars in which the magnetic field is much stronger. Similarly, we have neglected the turbulent $\alpha$ effect which (in comparison to the BL $\alpha$) is probably insignificant in the Sun but may be important in rapidly rotating stars \citep{Karak19}. Furthermore, we have not considered the small-scale magnetic field which could be generated through the small-scale dynamo and may become important in giving some non-thermal emission in the chromospheres of rapidly rotating stars. Finally, we have not considered any possible sources of the variations in the magnetic field (e.g., fluctuations in the BL process due to variations in the tilt angle and the variation in the meridional circulations). Observations 
show that many stars show irregular cycles and possibly extended grand minima \citep{Saar12,Wright16}. Thus the irregular aspects should be carefully studied, and checked whether the solar dynamo models \citep{KarakChou11,HKBC15,Kit18} which are successful in explaining many irregular aspects of the solar cycle are capable of explaining the irregular features of other stars. However, seeing the success in reproducing many observational features of stellar cycles in our model, we can have some confidence on the basic assumptions made in our model, and we expect that the prediction made on the quadrupolar field in rapidly rotating stars is true. We hope that future stellar observations will validate our prediction.

\begin{acknowledgements}
We thank an anonymous referee for constructive comments that helped to improve the manuscript. It is our pleasure to thank Prof. Arnab Rai Choudhuri for various discussions which helped us a lot to work on this project. GH thanks Prof. Aline Vidotto for several discussions related to this work. GH and JJ acknowledge the support by the National Science Foundation of China (grant Nos. 11873023, 11522325) and by the Fundamental Research Funds for the Central Universities of China. BBK sincerely thanks SERB/DST, India for providing research grand through the Ramanujan Fellowship (project no SB/S2/RJN-017/2018). BBK further thanks Janardhan Padmanabhan and Aveek Sarkar for providing the warm hospitality in Physical Research Laboratory, Ahmedabad, where he did some revision of this paper. LK acknowledges the support by the Russian Foundation for Basic Research (project 17-52-80064) and by budgetary funding of Basic Research program II.16.
\end{acknowledgements}

\bibliography{myref}
\end{document}